\RequirePackage{fix-cm}

\documentclass[twocolumn,epjc3]{svjour3_arxiv} 

\smartqed  

\RequirePackage{graphicx}

\RequirePackage{siunitx}
\RequirePackage[version=4]{mhchem}
\RequirePackage{booktabs}
\RequirePackage{multirow}
\RequirePackage[colorlinks,citecolor=blue,urlcolor=blue,linkcolor=blue]{hyperref}
\RequirePackage{dblfloatfix}
\RequirePackage{microtype}

\RequirePackage{amsmath}
\RequirePackage{amssymb}

\newcommand{\nue}{\nu_e}
\newcommand{\nuebar}{\overline{\nu}_e}

\newcommand{\avg}[1]{\left< #1 \right>}

\DeclareSIUnit\solarmass{\ensuremath{M_\odot}}
\DeclareSIUnit\parsec{\ensuremath{pc}}
\DeclareSIUnit\erg{\ensuremath{erg}}

\newcommand{\secref}[1]{Section~\ref{#1}}
\newcommand{\figref}[1]{Figure~\ref{#1}}
\newcommand{\litcite}[1]{Ref.~\cite{#1}}

\journalname{Eur. Phys. J. C}

\begin{document}

\title{Implementation and first results of the KM3NeT real-time core-collapse supernova neutrino search
}
\subtitle{The KM3NeT Collaboration}


\onecolumn

\begin{NoHyper} 



\author{
S.~Aiello\thanksref{a}
\and
A.~Albert\thanksref{ba,b}
\and
M.~Alshamsi\thanksref{c}
\and
S. Alves Garre\thanksref{d}
\and
Z.~Aly\thanksref{e}
\and
A. Ambrosone\thanksref{f,g}
\and
F.~Ameli\thanksref{h}
\and
M.~Andre\thanksref{i}
\and
G.~Androulakis\thanksref{j, dec}
\and
M.~Anghinolfi\thanksref{k}
% ...
}
\institute{
\label{a}INFN, Sezione di Catania, Via Santa Sofia 64, Catania, 95123 Italy
\and
\label{b}Universit{\'e}~de~Strasbourg,~CNRS,~IPHC~UMR~7178,~F-67000~Strasbourg,~France
\and
\label{c}Universit{\'e} de Paris, CNRS, Astroparticule et Cosmologie, F-75013 Paris, France
\and
\label{d}IFIC - Instituto de F{\'\i}sica Corpuscular (CSIC - Universitat de Val{\`e}ncia), c/Catedr{\'a}tico Jos{\'e} Beltr{\'a}n, 2, 46980 Paterna, Valencia, Spain
\and
\label{e}Aix~Marseille~Univ,~CNRS/IN2P3,~CPPM,~Marseille,~France
\and
\label{f}INFN, Sezione di Napoli, Complesso Universitario di Monte S. Angelo, Via Cintia ed. G, Napoli, 80126 Italy
\and
\label{g}Universit{\`a} di Napoli ``Federico II'', Dip. Scienze Fisiche ``E. Pancini'', Complesso Universitario di Monte S. Angelo, Via Cintia ed. G, Napoli, 80126 Italy
\and
\label{h}INFN, Sezione di Roma, Piazzale Aldo Moro 2, Roma, 00185 Italy
\and
\label{i}Universitat Polit{\`e}cnica de Catalunya, Laboratori d'Aplicacions Bioac{\'u}stiques, Centre Tecnol{\`o}gic de Vilanova i la Geltr{\'u}, Avda. Rambla Exposici{\'o}, s/n, Vilanova i la Geltr{\'u}, 08800 Spain
\and
\label{j}NCSR Demokritos, Institute of Nuclear and Particle Physics, Ag. Paraskevi Attikis, Athens, 15310 Greece
\and
\label{k}INFN, Sezione di Genova, Via Dodecaneso 33, Genova, 16146 Italy
% ...
\and
\label{ba}Universit{\'e} de Haute Alsace, rue des Fr{\`e}res Lumi{\`e}re, 68093 Mulhouse Cedex, France
% ...
}
\thankstext{corr1}{corresponding author, e-mail: km3net-pc@km3net.de}
\thankstext{dec}{deceased}
\thankstext{unipd}{also at Dipartimento di Fisica, INFN Sezione di Padova and Universit\`a di Padova, I-35131, Padova, Italy}

\date{}

\maketitle

\end{NoHyper} 

\twocolumn


\begin{abstract}
  The KM3NeT research infrastructure is under construction in the Mediterranean Sea. KM3NeT will study atmospheric and astrophysical neutrinos with two multi-purpose neutrino detectors, ARCA and ORCA, primarily aimed at GeV--PeV neutrinos. Thanks to the multi-photomultiplier tube design of the digital optical modules, KM3NeT is capable of detecting the neutrino burst from a Galactic or near-Galactic core-collapse supernova. This potential is already exploitable with the first detection units deployed in the sea. This paper describes the real-time implementation of the supernova neutrino search, operating on the two KM3NeT detectors since the first months of 2019. A quasi-online astronomy analysis is introduced to study the time profile of the detected neutrinos for especially significant events. The mechanism of generation and distribution of alerts, as well as the integration into the SNEWS and SNEWS 2.0 global alert systems are described. The approach for the follow-up of external alerts with a search for a neutrino excess in the archival data is defined. Finally, an overview of the current detector capabilities and a report after the first two years of operation are given.

\end{abstract}

\section{Introduction}
\label{s:intro}

The field of time-domain astronomy has seen considerable development in recent years due to its ability to study the extreme physics involved in cataclysmic phenomena like exploding stars, the birth of stellar black holes or the mergers of neutron stars. These sources can release enormous amounts of energy both as electromagnetic radiation and in non-electromagnetic forms such as neutrinos and gravitational waves (GW).

A core-collapse supernova (CCSN) is an explosive phenomenon that may occur at the end of the life of massive stars,  in which 99\% of the released energy is emitted in the form of quasi-thermal neutrinos in the $\SI{10}{MeV}$ energy range \cite{Giunti:2007ry}.

Neutrinos, gravitational waves and electromagnetic radiation are produced in the different phases of the CCSN evolution~\cite{MM_CCSN}, opening the possibility for multi-messenger observations. Neutrinos carry crucial information on the explosion mechanics and predate the optical observability of CCSNe acting as an early warning for a prompt optical follow-up. Gravitational-wave observatories may detect burst signatures from a CCSN simultaneously with its neutrino signal.
 
The only detection of CCSN neutrinos to date is from SN 1987A~\cite{SN1987A}, the explosion of a blue supergiant in the Large Magellanic Cloud at a distance of $\sim \SI{50}{\kilo\parsec}$. Two dozen neutrinos were observed by three neutrino detectors operating at that time: Kamiokande-II ~\cite{Kamioka}, IMB~\cite{IMB} and Baksan~\cite{Baksan}. This combined detection allowed to partially unveil the physical mechanism driving these explosions, showing that neutrinos play a major role in it~\cite{Janka}. However, the understanding of the core collapse dynamics is still incomplete~\cite{3Dreview}. 

Despite its huge flux magnitude, a CCSN neutrino burst is detectable only for Galactic and near-Galactic events due to the tiny interaction cross section at this energy. The expected rate of CCSNe in the Galaxy is estimated at $\sim 1.6 \pm 0.5$ per century~\cite{Rozwadowska:2021lll}. As the number and size of detectors sensitive to CCSN neutrinos grow, the collection of a large data sample, from $\mathcal{O}(\num{E3})$ to $\mathcal{O}(\num{E5})$ detected neutrinos, is expected from a future CCSN observation.

 While SN 1987A stands as one of the most remarkable multi-messenger observations to date, the neutrinos were found in the archival data of the detectors only after the optical discovery. Given the fundamental importance of the next CCSN observation, and since neutrinos can act as an early warning for optical observatories, the \emph{Supernova Neutrino Early Warning System} (SNEWS) \cite{Antonioli:2004zb} has been established to exploit the combined potential of the neutrino detectors in operation around the globe. The design principle followed by SNEWS aims at the dissemination of a \emph{prompt} and \emph{positive} alert, based on the coincident reporting of a CCSN detection by two or more experiments in the network within ten seconds. With this criterion, SNEWS achieves a false alarm rate (FAR) below one per century for its public alerts. As described further in this paper, KM3NeT is now part of the SNEWS network, contributing to the alert formation process by sending alerts with a rate below the required maximum FAR of one per week.

 Since the inception of the original SNEWS system, observatories have evolved their real-time follow-up capabilities into more advanced and flexible designs, being able to receive and filter a large amount of alerts from different providers. At the same time, multi-purpose alert networks have been established to coordinate the information exchange within the astronomical community. A notable example is the \emph{Gamma-ray Coordinates Network} (GCN), originally introduced to disseminate gamma-ray burst alerts produced by satellite-mounted instruments\footnote{\url{https://gcn.gsfc.nasa.gov}}, and grown to a general multi-messenger alert exchange system.

 Moreover, the capabilities of current and near-future neutrino detectors are largely enhanced compared to the past. In view of these developments, the original SNEWS network is being upgraded to a new generation system named SNEWS 2.0, with a multi-messenger program that aims to obtain the largest amount of information from the observation of the next supernova explosion \cite{SNEWS2.0}. The detailed study of the CCSN neutrino signature combined with a wide multi-messenger follow-up campaign will be the key to achieve this goal. For this, an early alert and a precise timing of the signal are crucial. 
 
 SNEWS 2.0 will introduce important new features on top of the existing SNEWS functionality. The coincidence-based mechanism for the alert generation will be augmented with a real-time significance-based combination of the detector signals. In addition, SNEWS 2.0
 allows for coordinated time-domain studies of the neutrino emission, enabling a variety of astrophysical analyses. Following a successful detection, the study of the time profile of the neutrino signal enabled by the combined effort of a new generation of neutrino detectors will be fundamental to understand the CCSN phenomenology. In a multi-messenger perspective, the precise timing of the signal is critical for the identification of a potential gravitational-wave counterpart, since providing a precise time window to gravitational-wave interferometers can enhance their detection sensitivity \cite{MM_CCSN}. Vice versa, a search for a neutrino counterpart can be performed as a follow-up of candidate gravitational-wave alerts. Furthermore, the detected neutrino light-curves can be combined to estimate the relative delays between the arrival times at each detector, allowing for the triangulation of the source location~\cite{Coleiro:2020vyj}.

 The KM3NeT Collaboration is building a deep-sea infrastructure in the Mediterranean Sea~\cite{KM3NeT:2016-LoI} hosting two Cherenkov neutrino detectors: a $\si{\kilo\metre^3}$-scale detector aimed at energies starting at the TeV scale (ARCA, \emph{Astroparticle Research with Cosmics in the Abyss}) and a \SIrange{6}{7}{\mega ton} densely instrumented detector for the lower neutrino energies down to the GeV scale (ORCA, \emph{Oscillation Research with Cosmics in the Abyss}). ARCA is located at about 100~km offshore Portopalo di Capo Passero in the South of Italy at a depth of 3500~m, and ORCA at 40~km offshore Toulon in the South of France, at a depth of 2450~m. The key element of the KM3NeT detectors is the Digital Optical Module (DOM), hosting thirty-one, $\SI{80}{\milli\metre}$ diameter, photomultiplier tubes (PMTs) enclosed in a pressure resistant glass sphere. DOMs are connected in groups of eighteen to form vertical detection units (DUs) which are anchored to the seabed. Each KM3NeT detector is built as an array of DUs connected to the shore by an underwater electro-optical cable for data and power transmission. A group of 115 DUs constitutes a building block. ARCA and ORCA will consist of two and one building blocks, respectively. Since both detectors are operated continuously from the installation of the first DUs, KM3NeT already aims for discoveries during its construction phase.

 In a KM3NeT detector, the identification and reconstruction of individual neutrinos are possible only for spatially extended or sufficiently bright Cherenkov signatures, typical of interactions above a threshold of a few GeVs. However, neutrinos from a CCSN burst interacting in the vicinity of the KM3NeT DOMs can be detected as a global increase in the coincidence rate of the photons recorded by the optical modules. This method enables KM3NeT to detect a Galactic or near-Galactic supernova, as detailed in Ref.~\cite{KM3NeT2021-CCSN}. CCSN neutrinos are primarily detected in water through the inverse beta decay process ($\nuebar$) with secondary contributions from elastic scattering on electrons (all flavours) and oxygen interactions ($\nue$ and $\nuebar$)~\cite{Scholberg:2012id}. 

 This paper illustrates the real-time strategy for the detection and follow-up of CCSN events in KM3NeT The real-time realisation of the detection method introduced in Ref.~\cite{KM3NeT2021-CCSN}, followed by the implementation of a CCSN trigger combining the data from the ARCA and ORCA detectors, are described in Section~\ref{s:pipeline}. The trigger routinely generates internal alerts with different levels of significance, which are distributed to the KM3NeT multi-messenger infrastructure. As a response to a high-significance detection, a quasi-online astronomy analysis is applied to purposely buffered low-level data, in order to determine the time profile of the neutrino signal and estimate the arrival time of the burst. The buffered data are then stored for offline studies, such as the estimation of the neutrino spectral parameters and the search for the signature of hydrodynamical instabilities, as further detailed in Ref.~\cite{KM3NeT2021-CCSN}. The workings of the quasi-online analysis are introduced in Section~\ref{s:quasi-online}. The overall alert generation strategy is detailed in Section~\ref{s:alerts}. Thanks to its CCSN trigger, KM3NeT is currently able to send alerts to SNEWS, with the potential of sharing with SNEWS 2.0 the results of the quasi-online analysis. A strategy for the follow-up of external alerts with a triggered search in archival data is outlined in Section~\ref{s:triggered_search}. The aim of the triggered search is to provide a detection significance or upper limits for the observation of a low-energy neutrino signal corresponding to a specific alert time. In Section~\ref{s:current}, the current status of the CCSN online search system and the results of the first astrophysical follow-ups of externals triggers are summarised. Conclusions and outlooks are drawn in Section~\ref{s:conclusion}.
 
 The benchmark CCSN neutrino fluxes referenced in this work come from three-dimensional simulations of three progenitors, with stellar masses of $\SI{11}{\solarmass}$, $\SI{27}{\solarmass}$ and $\SI{40}{\solarmass}$, developed by the MPA Garching group \cite{Tamborra:2014hga-3D,Walk:2019miz-3DBH}. The corresponding particle fluences have a quasi-thermal spectrum with mean energies of about 13.7, 15.7 and $\SI{18.2}{MeV}$, respectively. The $\SI{27}{\solarmass}$ CCSN has roughly twice the $\nuebar$ luminosity than the $\SI{11}{\solarmass}$. The $\SI{40}{\solarmass}$ progenitor collapses into a black-hole, resulting in a failed supernova characterised by a strong neutrino emission, almost four times the $\SI{11}{\solarmass}$ case, without optical counterpart. Fluence plots for the three scenarios are given in Ref~\cite{KM3NeT2021-CCSN}.

\section{Real-time CCSN search pipeline}
\label{s:pipeline}

 The approach used to detect neutrinos from a core-collapse supernova burst in KM3NeT has been detailed in Ref.~\cite{KM3NeT2021-CCSN}. In this section, the implementation of such a method in the form of a real-time processing pipeline is described. The primary goal of the pipeline is to analyse the data collected by both KM3NeT detectors in order to evaluate a time-dependent significance that can be exploited to identify a CCSN signal and generate alerts, or be shared in real time with an external network such as SNEWS 2.0~\cite{SNEWS2.0}.

\begin{figure*}
    \centering
    \includegraphics[width=0.70\textwidth]{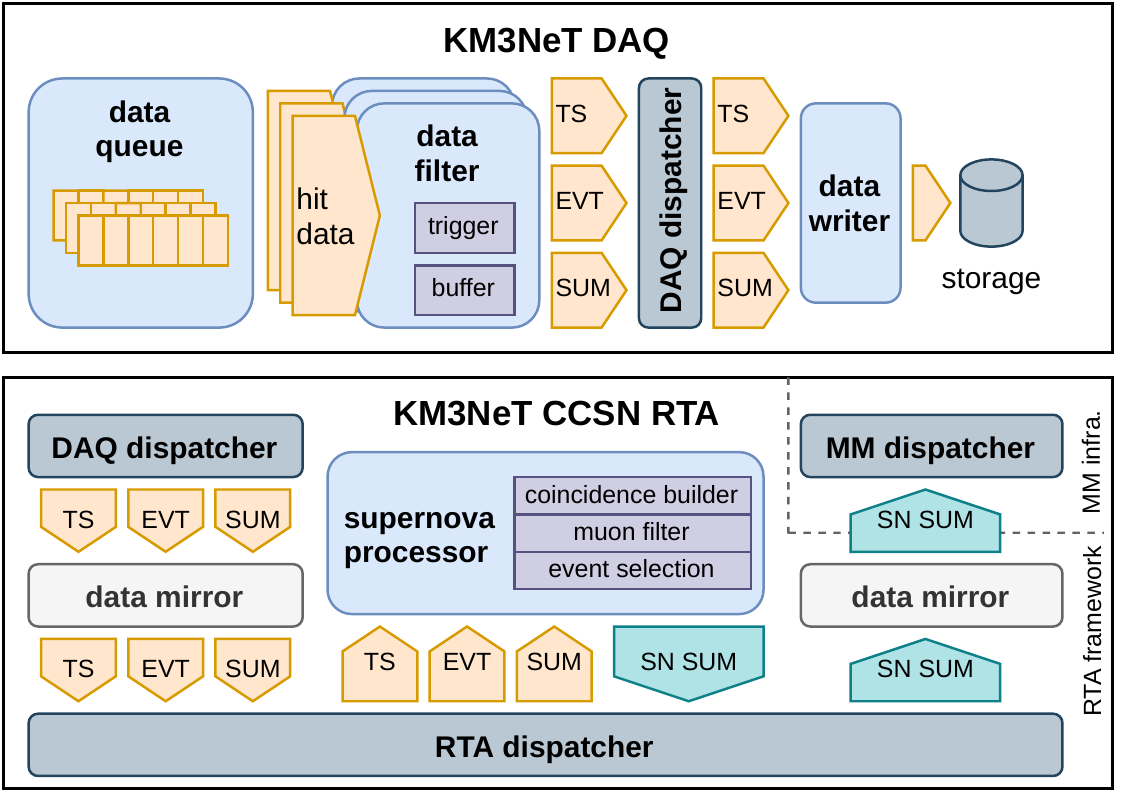}
    \caption{Logic diagram of the data acquisition system of KM3NeT detectors (top) and the CCSN real-time analysis (RTA) framework (bottom) running at the computing farm of each detector's shore station. The data streams are indicated with the following abbreviations: TS, timeslices; EVT, triggered events; SUM, summaryslices (PMT rates); SN SUM, supernova summary.
    The dashed line separates the local processing framework from the common KM3NeT multi-messenger infrastructure.}
    \label{fig:rta-DAQ}
\end{figure*}

 The pipeline is articulated in two stages. The first is the \emph{real time analysis} (RTA) framework local to the computing farm of each detector's shore station and integrated with the corresponding data acquisition system (see Section~\ref{s:daq}). At this stage, the detector data are analysed to produce a \emph{supernova summary} (SN SUM). The supernova summary is a message containing the number of events identified by the analysis algorithm for the supernova search (\emph{SN events}, as defined in Section~\ref{s:rta}). A supernova summary is produced every $\SI{100}{\milli\second}$ for the time window covering the previous $\SI{500}{\milli\second}$. Auxiliary information is included for quality control and monitoring of the detector status.  An overview of the KM3NeT data acquisition system \cite{Real:2015lwv-DAQ,Bruijn:2020olw-DAQ,Pellegrino:2016cas-DAQ} and the CCSN real-time analysis framework is given in \figref{fig:rta-DAQ}.

 In the second stage of the real-time chain, a joint time-dependent supernova search significance (SN Z) is evaluated by the KM3NeT CCSN trigger, combining the data from the ARCA and ORCA detectors. The trigger is part of the common KM3NeT multi-messenger infrastructure hosted at a dedicated computing farm. The significance information is exploited by the alert generation system as described in \secref{s:alerts}.

 The two processing stages are supported by three instances of a data dispatcher software, designated as the ARCA/ORCA \emph{DAQ dispatchers} and the \emph{KM3NeT multi-messenger dispatcher} respectively. The supernova summary data of each detector are forwarded from their respective DAQ dispatcher to the KM3NeT multi-messenger dispatcher. The data exchange protocol allows for the transmission of \emph{tagged} messages containing either text or binary data. A \emph{tag} identifies a single data stream. Different clients can connect in order to send and/or receive one or more data streams to/from the dispatcher, with a tag-based selection. Text messages exchanged through the dispatcher for the scope of this analysis are formatted according to the JavaScript Object Notation (JSON)\footnote{\url{https://json.org}} standard.

\subsection{Data acquisition in KM3NeT}
 \label{s:daq}
 The KM3NeT detectors adopt an \emph{all-data-to-shore} data acquisition concept. The data  acquisition system 
 is outlined in \figref{fig:rta-DAQ} (top). In the KM3NeT DOM, a \emph{hit} is recorded whenever the signal of any PMT passes a threshold of 0.3 photoelectrons. The time of the leading edge of the signal and the time-over-threshold are recorded with nanosecond precision by the front-end electronics enclosed in the DOM \cite{Aiello:2019qol-CLB,KM3NeT:2021yiw-FW}. The corresponding data are packed and sent to shore in segments of $\SI{100}{\milli\second}$. A high rate veto logic is implemented to discard the data from any PMT recording more than 2000 hits over a $\SI{100}{\milli\second}$ segment. Bioluminescence in the sea is the typical origin of such occasional high rates. A DOM remains active even when one or more PMTs are disabled. At the computing farm onshore, the data are collected by a software \emph{data queue} and distributed to an array of software data filter nodes to be further processed. The data filter software applies different trigger algorithms to identify candidate events due to high-energy (above $\sim \SI{1}{GeV}$) neutrino interactions occurring inside or in the vicinity of the detector. The first operation of the data filter is the identification of coincidences consisting of two or more hit PMTs on the same DOM, within a time window of $\SI{10}{\nano\second}$. Then, clusters of causally connected coincidences and single hits occurring within a cylindrical or spherical volume are searched for in order to identify track-like signatures from muons or shower-like signatures from electromagnetic cascades, respectively. The discrimination of the atmospheric muon background is performed in the later analysis of the stored data.

 Three main types of data are produced as output of the data filters: \emph{triggered events} (EVT), \emph{timeslice data} (TS) and \emph{summaryslice data} (SUM).
 
 A triggered event contains information on the set of causally connected hits selected by the trigger algorithm as part of an event topology of interest (\emph{triggered hits}). The triggered hits are stored in a so-called \emph{snapshot} together with all the background hits recorded in a time interval extending before and after the triggered hits by a designated margin.

 Timeslice data consist of the hit data collected from all the DOMs in a $\SI{100}{\milli\second}$ time interval, and selected according to coincidence criteria based on hit time difference, multiplicity and angular separation between PMTs.
 Several types of timeslices are produced according to different selections, oriented to specific analysis purposes.
 
 For the supernova search, the \emph{SN timeslice} stream considers all hits selected on the basis of coincidences with at least four hit PMTs in a DOM whose axes are within 90 degrees, in a time window of $\SI{15}{\nano\second}$. Below this multiplicity, the signal to noise ratio is unfavourable to the identification of a signal, since backgrounds are largely predominant over the coincidence rates due to interactions of low-energy neutrinos from a CCSN \cite{KM3NeT2021-CCSN}. The width of the time window used by the data filter, $\SI{15}{\nano\second}$, is slightly larger than physically required. A shorter time interval (e.g. $\SI{10}{\nano\second}$) is adopted in the later analysis, where a more accurate time calibration can be applied. The coincidence rate recorded in SN timeslices is about $\SI{5}{\hertz}$ per DOM, with a data throughput of $\SI{250}{kB.s^{-1}}$ per building block.
 
 In parallel, the \emph{L1 timeslice} stream containing all hits in coincidence within a $\SI{25}{\nano\second}$ time window is produced for detector calibration. With an average coincidence rate of about $\SI{1}{kHz}$ per DOM, the L1 timeslice data throughput of a KM3NeT building block is of the order of $\SI{24}{MB.s^{-1}}$. Since the permanent storage of the whole L1 data is not sustainable, the data filter performs a downsampling of this timelice stream according to a configurable decimation factor, $s$, meaning that only one every $s$-th timeslice is kept at the output. To allow the use of these data in astrophysical follow-ups, timeslices are temporarily stored on a local storage space by each instance of the data filter application, and dumped on request, as further described in \secref{s:quasi-online}.
 
 The summaryslice data consist of the individual PMT rates, derived from the raw hit data and averaged over the corresponding $\SI{100}{\milli\second}$ of a timeslice, plus status information for each PMT acquisition channel. These data are in general exploited for monitoring and are auxiliary to calibration.

 After the filtering and triggering operations, the data filters send their output data to the DAQ dispatcher. From this, a \emph{data writer} application writes all the data produced by the data filters on local storage. In parallel, the same data streams can be processed by different real-time analysis applications. 
 
 The data acquisition system is driven by the Control Unit software, as described in Refs.~\cite{KM3NeT:2019hqg-CU,Bozza:2019osq-CU}.

\subsection{Real-time data analysis}
\label{s:rta}
 The first stage in the real-time supernova search is the analysis of the hit data in SN timeslices, aimed at evaluating the number of \emph{SN events} as a function of time. An SN event is defined as a coincidence on a DOM that matches the selection criteria of the supernova search. This processing stage is implemented in an application referred to as \emph{supernova processor} (see \figref{fig:rta-DAQ}, bottom).
 
 Since timeslices are produced by an array of data filters running in parallel, their order of arrival at the input of the supernova processor is not guaranteed. The incoming timeslices are buffered in a priority queue in order to be time sorted before being processed. The length of the priority queue is by default set to 100 timeslices, therefore spanning 10 seconds, but can be optimised in order to reduce the latency, if required. Triggered events and summaryslices are also collected by the supernova processor to be exploited in the analysis.

 Following the approach described in Ref.~\cite{KM3NeT2021-CCSN}, the analysis of SN timeslice data consists of three steps:
 \begin{enumerate}
    \item \emph{Coincidence building}. The hits in SN timeslices are selected by the data filter and stored without preserving any information on the original coincidences. For this reason, the supernova processor applies again a coincidence building algorithm on the hit data, searching for coincidences among different PMTs on the same DOM, throughout all the DOMs of the detector. A coincidence at the DOM level is defined as the detection of two or more hits within a $\SI{10}{ns}$ time window starting at the first hit. All coincidences found according to this criterion in the 100 ms of the SN timeslice are considered for the next step of the analysis.
    \item \emph{Background reduction and filtering}. The first purpose of this stage is to effectively reduce to a single entity a group of multiple coincidences having the same physical origin, such as in the case of atmospheric muons (multiple coincidences on different DOMs) and PMT afterpulses (multiple coincidences on the same DOM).
    Clusters are identified as sequences of coincidences occurring within a fixed time window. The time window, chosen to fit the typical crossing time of a muon going through the instrumented volume, is of $\SI{1}{\micro\second}$ for the ORCA detector and $\SI{3}{\micro\second}$ for the ARCA detector. For each cluster, only the coincidence with the highest multiplicity is considered. If more than one coincidence has the same highest multiplicity, only the earliest is kept. The remaining coincidences in the cluster are discarded, and the search is continued starting at the end of the time window. After the reduction, a filter to suppress the contribution of atmospheric muons is applied. In this, the data from the triggered events occurring in the time spanned by the SN timeslice are used to define a \emph{veto} on all the DOMs that have detected at least one triggered hit. Coincidences occurring on these DOMs are discarded, as they are likely coming from an atmospheric muon crossing the instrumented volume.
    
    \item \emph{Event selection}. Coincidences in a given multiplicity interval are selected, and defined as \emph{SN events}. Such interval is chosen to maximise the distance to the source at which a $5\,\sigma$ significance can be obtained for the detection of the flux corresponding to the \SI{11}{\solarmass} progenitor. For the complete KM3NeT detector (three building blocks), the optimal multiplicity selection has been estimated to be 7--11 \cite{KM3NeT2021-CCSN}. To compensate for the lower statistics of a partial detector installation, the real-time search operating in the current configuration of KM3NeT (twelve detection units in total) adopts a wider selection interval of 6--11. The coincidences beyond multiplicity 11 are excluded to increase the robustness of the selection, given that their contribution to the significance is negligible \cite{KM3NeT2021-CCSN}.
 \end{enumerate}   

 From the analysis of the detector data, the supernova processor builds a \emph{supernova summary} (SN SUM) containing: (i) the starting time of the $\SI{500}{\milli\second}$ interval, (ii) the number of SN events in the interval, (iii) the number of DOMs detecting at least one SN event in the interval, (iv) the number of active DOMs in the detector, (v) the number of active PMTs in the detector, and (vi) the total hit rate of the detector estimated from the summaryslice data. This information is propagated to the multi-messenger dispatcher, collecting the data coming from both the ARCA and ORCA detectors (see \figref{fig:realtime-system}).

\subsection{The KM3NeT CCSN trigger}
 \label{ss:trigger}
 The \emph{KM3NeT CCSN trigger} is a dedicated application that receives the information produced by the instances of the supernova processor running in the ARCA and ORCA shore stations. The role of the trigger is to compute the individual detection significances, and to combine them into a single KM3NeT significance. A quality score is assigned to the significance evaluation in order to discriminate against instrumental anomalies. This information is enclosed in a dedicated text message (SN Z) and reported to the multi-messenger dispatcher, in order to be processed by the alert sending and monitoring routines.

 \subsubsection{Data collection and synchronisation}
 
 The first task of the trigger application is the management of the incoming data. Since ARCA and ORCA are operated independently, the CCSN trigger is designed so that a downtime of one detector does not prevent the trigger from operating in stand-alone mode with the data coming from the other site. In addition, the trigger needs to tolerate variations in the latency of each detector data stream, defined as the difference between the time at which the hit data are generated offshore and the time at which the corresponding SN summary arrives at the dispatcher. To meet these requirements, the trigger application implements an adaptive queue for the incoming data. If the earliest SN summary in the queue corresponds to detector $A$ at time $t_i$, the application waits for the corresponding SN summary from detector $B$ for a fixed time interval, queuing further incoming data from detector $A$, before switching to single detector operation. In normal conditions, the storage of two SN SUM messages per detector ensures that the data are processed synchronously and with negligible added delay. The processing holds for a limited time (of the order of a few seconds) only when a detector stops sending data. The queue logic ensures that the combined operation is recovered at the earliest, as soon as the offline detector resumes the data transmission. While currently operating with two sites, the chosen implementation allows for the combination of an arbitrary number of detectors.
 
 \subsubsection{Quality score}
 Occasional anomalies in the KM3NeT DOM behaviour, such as the output of noisy or sparking PMTs, may result in a generation of spurious coincidences from a subpopulation of DOMs, or even from a single DOM. While these circumstances are rare and do not affect the general reliability of an offline search, they can seriously impact the real-time analysis by producing false signals that fall outside the statistical model for the background. For this reason, the consistency between the total number of SN events and the number of DOMs detecting at least one SN event is checked, and a quality score is assigned to the event count of each SN summary coming from a detector. For a given number of SN events, $n$, and a total number of DOMs in the detector, $m$, the probability distribution of the number of DOMs detecting at least one event, $k$, is:
 \begin{equation}
    \label{eq:qscore}
    p_{n,m}(k) = \frac{m!\,S(n, k)}{(m-k)!\,m^n} 
 \end{equation}
 where $S(n, k)$ is the Stirling partition number (also known as Stirling number of the second kind) for $n$ objects and $k$ subsets \cite{johnson1977urn,Kolchin-10.1137/1022018}. For an observation of $n$ SN events from $k$ DOMs, given that $m$ DOMs are active, the quality score is defined as the probability of observing $n$ events originating from a number of DOMs lower than or equal to $k$, and is derived from the cumulative distribution of Eq.~\eqref{eq:qscore}. A low quality score indicates that one or a few DOMs are producing an anomalous number of coincidences, a behaviour not compatible with the hypothesis of an evenly distributed background or signal population of events. A threshold on the quality score is implemented to veto the alert generation (see \secref{s:alerts}). Exploiting the data from long-term monitoring of the detector, the value of the threshold is chosen to ensure that all the instrumental anomalies identified by an irregular distribution of the SN events across the DOMs are rejected. After the quality selection, the distribution of the accepted significances is consistent with the Poisson expectation. The quality score threshold adopted in the current configuration is 0.001, resulting in a one-per-mille inefficiency in the regular alert generation.
  
 \subsubsection{Adaptive background expectation}
 \label{ss:adaptive}
\begin{figure}
    \centering
    \includegraphics[width=0.45\textwidth]{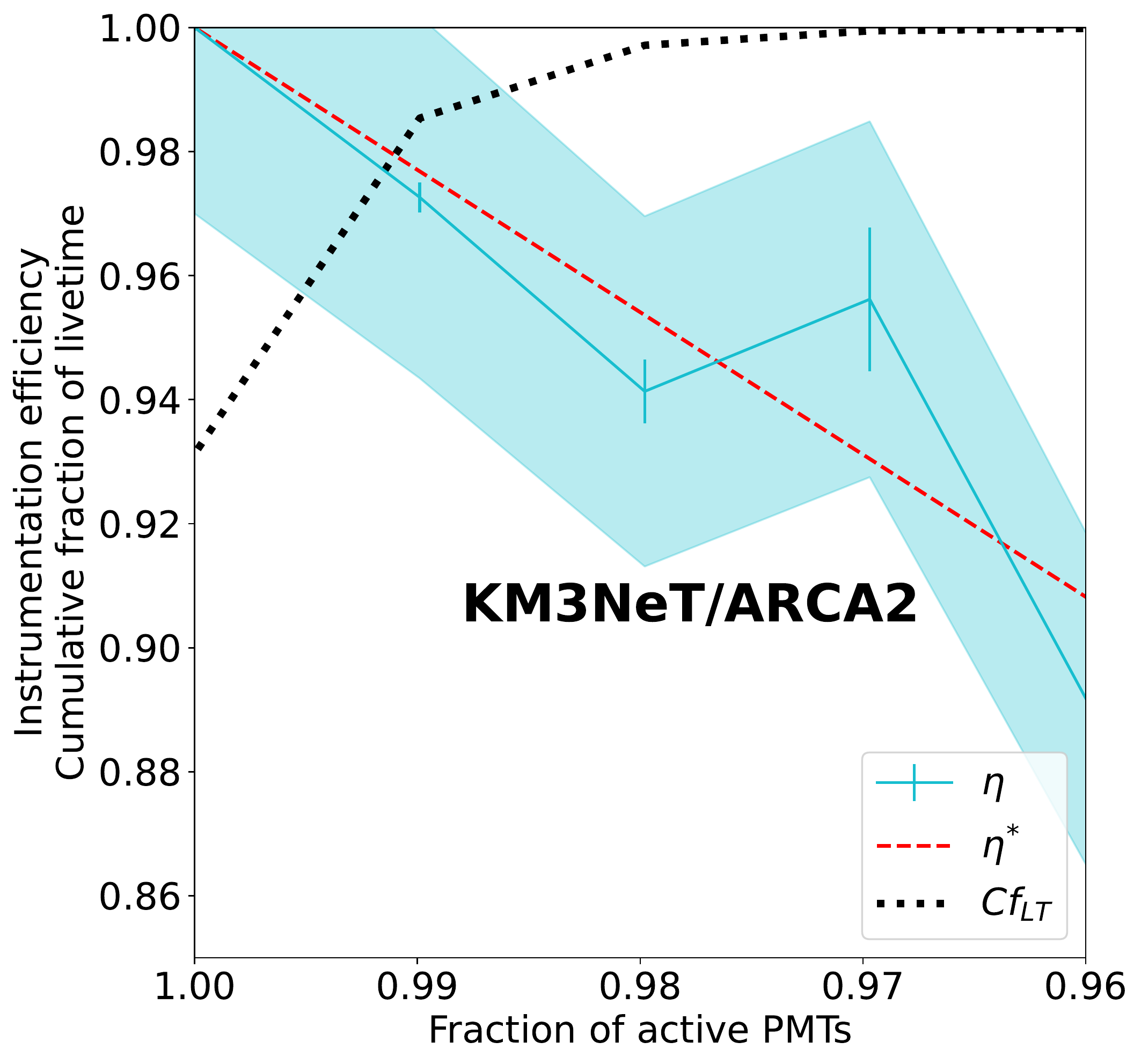}%
    \\
    \includegraphics[width=0.45\textwidth]{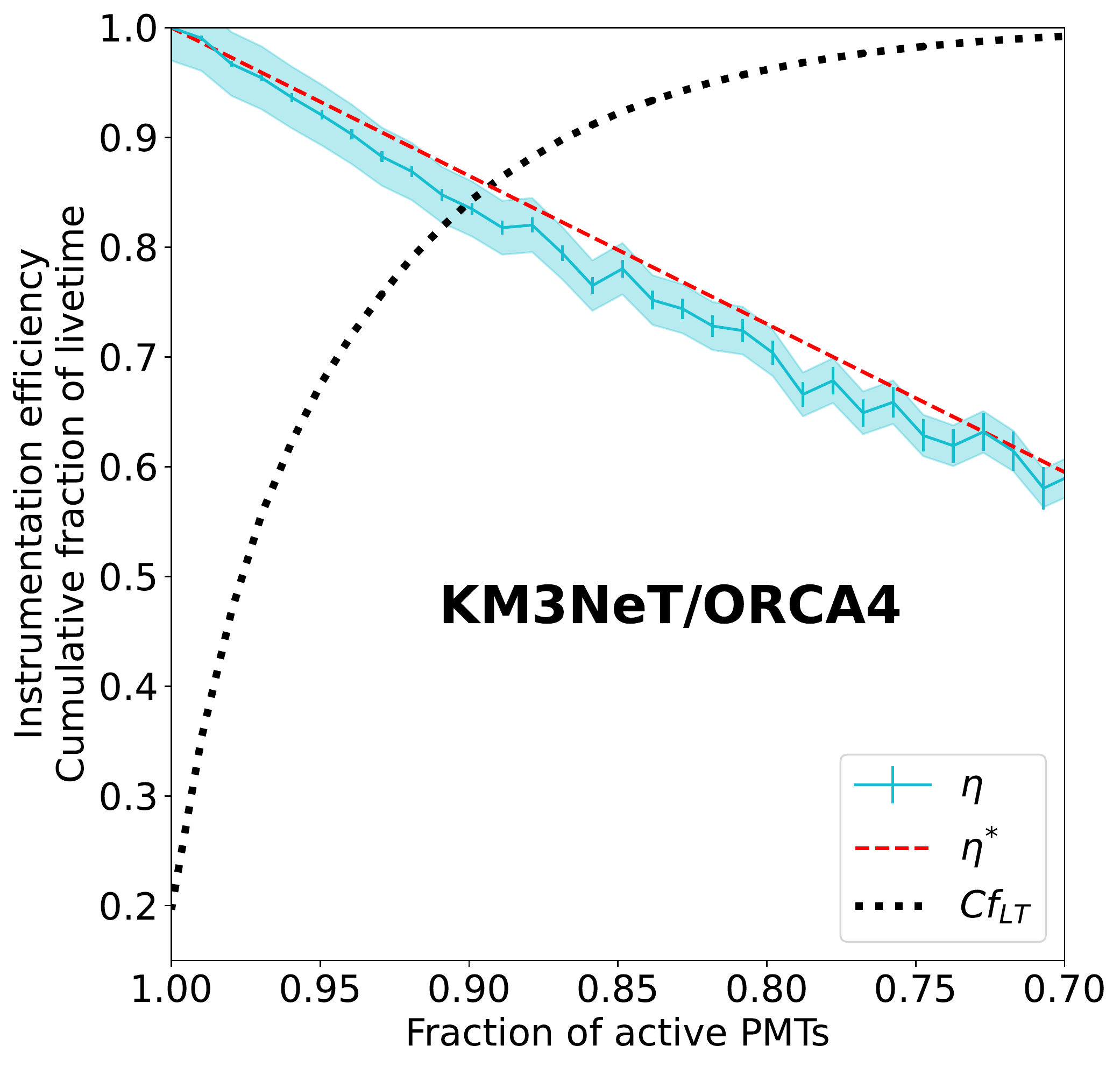}%
    \caption{Measured instrumentation efficiency, $\eta$, as a function of the fraction of active PMTs in the ARCA2 (top) and ORCA4 (bottom) detectors. Statistical errors are shown as error bars. Monthly variability is covered by the shaded area. The parameterisation, $\eta^{*}$, adopted to scale the background expectation in the trigger is drawn as a red dashed line. The cumulative fraction of livetime, $Cf_{LT}$, representing the fraction of livetime for which the detector has more than a given fraction of active PMTs, is superimposed as a black dotted line.}
  \label{fig:efficiency-parameterisation}
 \end{figure}

 Due to the variations of bioluminescence activity and the adoption of a high rate veto logic in the front-end electronics \cite{Aiello:2019qol-CLB}, the number of active PMTs is not constant in time and changes at each timeslice. For a correct evaluation of the significance, the expectation value for the number of background events needs to be known independently of the fraction of active PMTs. To this purpose, the \emph{instrumentation efficiency}, $\eta$, is defined as the ratio between the rate of background SN events measured with a given fraction of active PMTs, $f_A$, and the background rate measured with a fully active detector ($f_A \simeq 1$). The dependence of $\eta$ on $f_A$ is estimated through an offline analysis of archival data for a given detector configuration. This dependence is shown in \figref{fig:efficiency-parameterisation} for the ARCA detector with two detection units (ARCA2) and the ORCA detector with four detection units (ORCA4). The monthly variability of the instrumentation efficiency evaluated over a few-months time period, is within $\pm 3\%$. The two plots cover the range of values of the fraction of active PMTs for which a relevant amount of livetime is recorded. This is different between ARCA and ORCA due to the difference in the incidence of bioluminescence bursts at the two sites.
 
  In order to exploit this relation in the real-time calculation of the significance, a linear interpolation of the instrumentation efficiency, $\eta^{*}$, is adopted:
  \begin{equation}
     \eta^{*}(f_A) = 1 - \beta^{*} (1-f_A) \;
  \end{equation}
  where the $\beta^{*}$ coefficient represents the ratio between the assumed efficiency loss and the fraction of non-active (vetoed) PMTs. The value $\beta^{*}$ is estimated through a fit of the measured instrumentation efficiency, $\eta$, from the archival data of the ORCA4 and ARCA2 detectors. Conservatively, $\beta^{*}$ is determined by reducing the fit result of 10\% to prevent underestimations of the background as a consequence of the observed deviations from linearity. The resulting value of $\beta^{*}$ is of $\sim 2.3$ for ARCA and $\sim 1.35$ for ORCA. The parameterisation is drawn in \figref{fig:efficiency-parameterisation} as a red dashed line. As an additional safety measure, the value of $\eta$ for $f_A \simeq 0.7$ is assumed as a floor value for any lower fraction of active PMTs. In the real-time processing, $f_A$ is calculated from the number of active PMTs and the number of active DOMs, determined by the supernova processor through the analysis of summaryslice data. For the determination of the search significance (SN Z) corresponding to a \SI{500}{\milli\second} time window, the instrumentation efficiency is calculated for each 100~ms timeslice, and averaged. The background expectation value in the CCSN trigger is accordingly scaled, in order to provide a consistent estimate of the significance across the different data taking conditions. As new detection units are installed with the progress of the detector construction, the above parameterisation will be re-assessed on the basis of the new available data.
 
\subsubsection{Significance estimation}
 The number of background SN events in each interval of $\tau = \SI{500}{\milli\second}$ is distributed according to Poisson statistics, with an expectation value $\mu_b = m\,\rho\,\tau$, where $m$ is the number of active DOMs of the detector and $\rho$ is the background SN event rate per DOM. The reference value of $\mu_b$ is established from the offline analysis of a sufficiently large sample of timeslice data. A fixed value is provided as an input to the trigger application, and scaled as function of the fraction of active PMTs according to the logic described in the previous section.

 The one-sided p-value of the observation is calculated from Poisson statistics and converted into its equivalent Gaussian significance, $Z$. The significances as a function of the distance, $d$, evaluated individually for the ARCA and ORCA detectors are combined adopting a weighted linear combination:
 \begin{equation}
    \label{eq:combination}
    Z_{\textrm{KM3NeT}}(d) = \frac{\sum_{\alpha \in \{\textrm{ARCA}, \textrm{ORCA}\}} w_\alpha Z_\alpha (d)}{\sqrt{\sum_{\alpha \in \{\textrm{ARCA}, \textrm{ORCA}\}} w_\alpha^2}}
 \end{equation}
 where the weight, $w$, is chosen as the detection sensitivity at a reference distance of $\SI{10}{kpc}$, $w_\alpha = Z_\alpha(d = \SI{10}{kpc})$ for a benchmark flux model, chosen as the $\SI{27}{\solarmass}$ stellar progenitor. The choice of a different benchmark flux does not change the relative weights.
 
 From the combined significance, the false alarm rate is calculated as the product of the corresponding p-value and the update frequency of the search window, corresponding to $(\SI{100}{\milli\second})^{-1}$. This estimation of the FAR is conditioned by the fact that consecutive evaluations of the significance are based on overlapping time intervals, and a single fluctuation could be counted multiple times. This will be accounted for in the generation of alerts with the application of a dead time, as discussed in \secref{s:alerts}.

 The trigger application encapsulates in a message the relevant information: (i) combined significance, (ii) equivalent FAR, (iii) quality score, and (iv) number of SN events and DOMs for each detector. This information is reported to the common multi-messenger dispatcher in the form of a JSON message.

\section{Quasi-online astronomy analysis}
 \label{s:quasi-online}
 The CCSN detection in KM3NeT relies on a high-purity data sample based on high-multiplicity coincidences, which allow to largely reduce the contribution to the background by seawater radioactivity. This approach maximises the detection sensitivity and triggering horizon. At the same time, KM3NeT can collect a much larger event statistics by evaluating all the detected coincidences, at the cost of a lower signal-to-noise ratio. This  all-coincidence data sample is used to evaluate the time profile of the neutrino signal (detected neutrino light-curve) and investigate its features. The analysis of such sample is a fundamental part of the KM3NeT strategy for a CCSN neutrino observation, as outlined in Ref.~\cite{KM3NeT2021-CCSN}. 
 
 As introduced in \secref{s:rta}, the L1 timeslice stream collects all the hits belonging to coincidences on a $\SI{25}{\nano\second}$ scale. These data are routinely exploited for calibration purposes and monitoring of the detector performance \cite{KM3NeT:2019-MuonDepth}. Due to the applied downsampling, the L1 data that are routinely stored are not suitable for the follow-up of a transient candidate.
 
 In case of a significant detection, the recording of a continuous L1 data sample around the time of the event is required. To this purpose, a circular buffer is implemented in the data filter software to allow the continuous storage of a designated number of timeslices. The data are cached in a file residing on the local storage of the machine hosting the data filter instance. The buffer size is chosen to cover an interval of the order of 10 minutes. On request, the buffer is dumped, i.e. the file is saved in its current state and a new one is started. The dump of the buffer is requested with an internal alert, generated every time the trigger significance is above a defined threshold. The generation of internal alerts is further covered in \secref{s:alerts}. In the typical configuration, ten minutes of data centred on the alert time are stored in the buffer dump. As soon as the data are ready, a routine is launched to apply a coincidence analysis and determine the time profile of the signal.
 
 The coincidence time window is here decreased to $\SI{5}{\nano\second}$, to reduce the contribution of random correlation of independent photons~\cite{KM3NeT2021-CCSN}. The number of detected coincidences as a function of time is then evaluated with a resolution of 1~ms. The result of the analysis is stored in a \emph{light-curve summary} (LC SUM) message, consisting of a reference UTC start time and a vector of coincidence count values covering the time frame of the buffered data. The result obtained by the analysis performed at each detector is reported to the common multi-messenger dispatcher.
 
 \begin{figure*}
    \centering
    \includegraphics[width=0.65\textwidth]{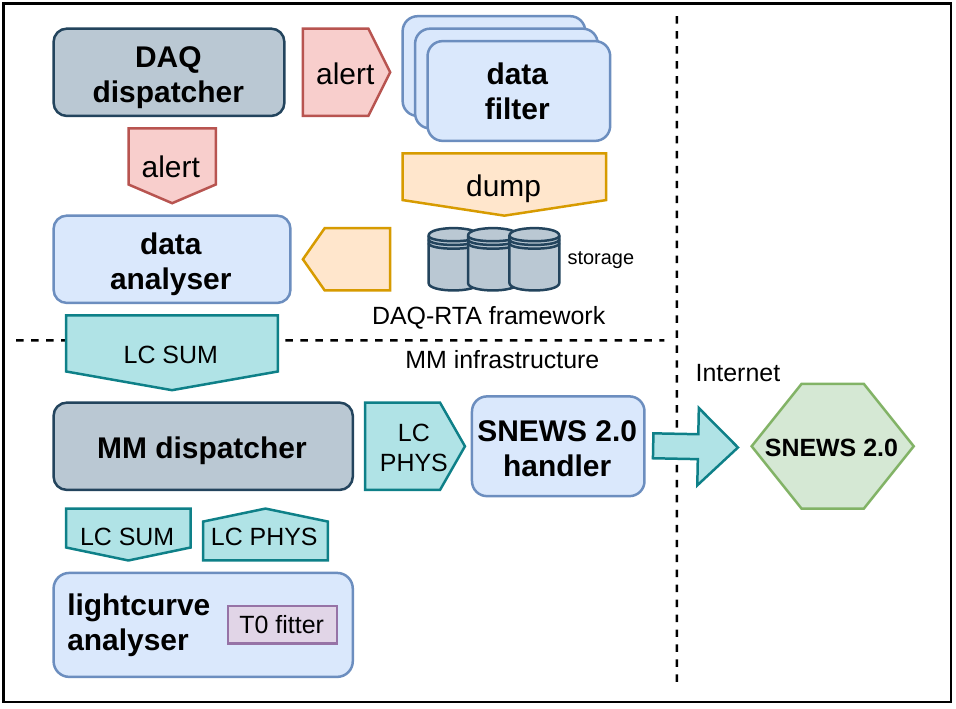}
    \caption{Overview scheme of the quasi-online analysis.}
    \label{fig:quasi-online}
\end{figure*}
 
 The analysis of the neutrino light-curve is carried out by the \emph{light-curve analyser} application with two main targets. The first is the precise determination of the signal arrival time through an exponential fit to the signal leading edge, as described in \litcite{KM3NeT2021-CCSN}. This is implemented in the \emph{T0 fitter} module. The second is the real-time sharing of the light-curve data with the SNEWS 2.0 network. Further analyses can be foreseen on the archived data. The light-curve analyser produces a \emph{light-curve physics} (LC PHYS) message that is distributed to the multi-messenger dispatcher and subsequently forwarded to the SNEWS 2.0 network by a dedicated application (\emph{SNEWS 2.0 handler}). In \figref{fig:quasi-online}, an overview of the quasi-online pipeline is given.

\section{Alert generation and dissemination}
\label{s:alerts}

\begin{figure*}
    \centering
    \includegraphics[width=0.70\textwidth]{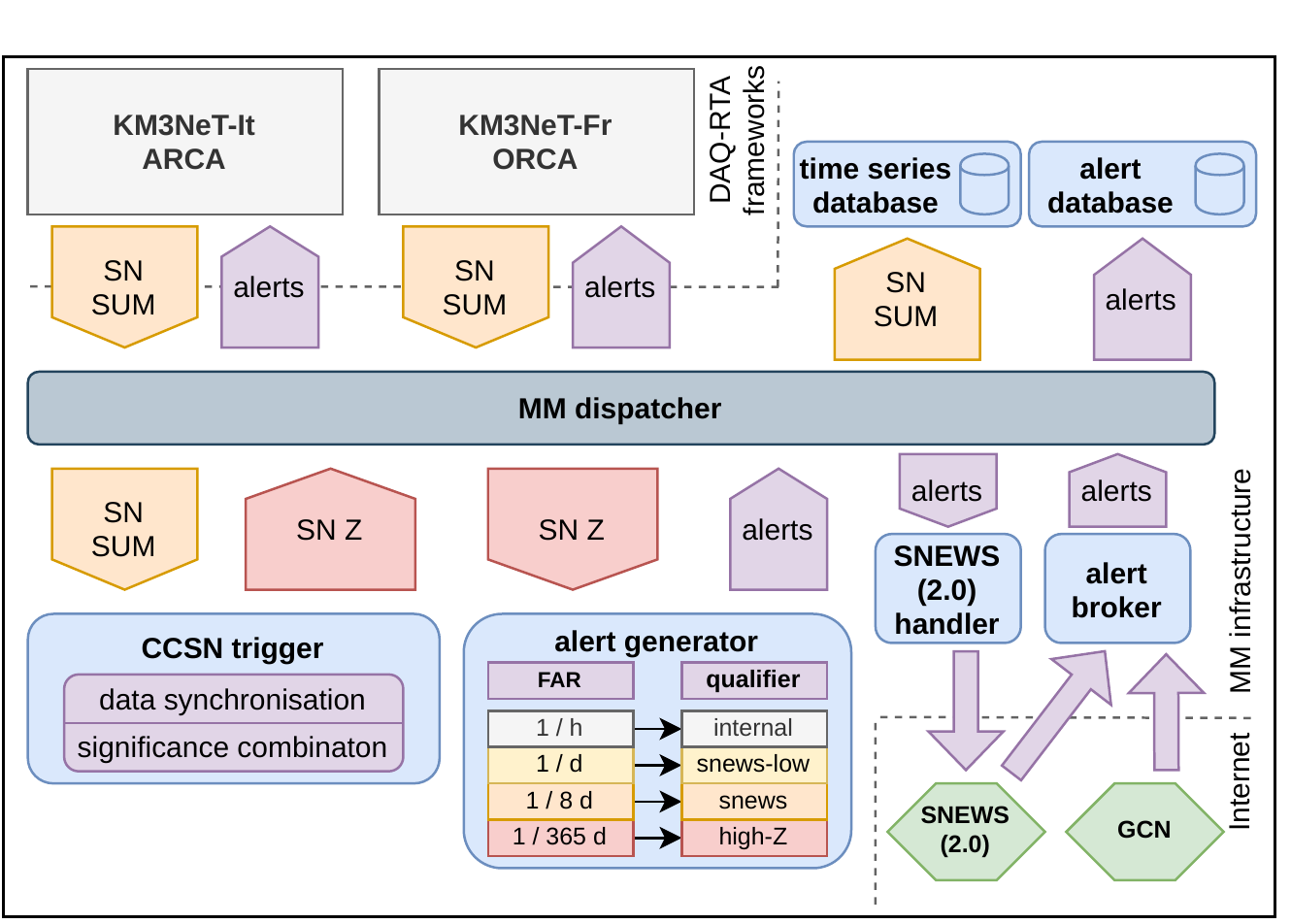}
    \caption{Overview of the CCSN alert generation and management data flow in the KM3NeT real-time system.}
    \label{fig:realtime-system}
\end{figure*}

The real-time analysis introduced above has the goal to continuously process the data streams from the KM3NeT detectors and estimate a time-dependent significance. An alert generation application is introduced to select interesting events based on given significance thresholds, in order to trigger the quasi-online analysis, and propagate the relevant alert information to external networks. This mechanism is outlined in \figref{fig:realtime-system}. In this view, an internal alert generator is configured with a list of thresholds expressed in terms of a maximum false alarm rate and a set of corresponding qualifiers indicating the type of alert to be generated. The alert generator receives the output of the CCSN trigger (see \secref{ss:trigger}) from the multi-messenger dispatcher. Each received message (SN Z) reports the significance and the corresponding FAR. The latter is compared with the issued thresholds, and the corresponding alert messages are generated and transmitted to the multi-messenger dispatcher for further processing. An internal alert message consists of a JSON dictionary and it includes a unique ID, the alert type (e.g. `supernova'), the qualifier, the UTC timestamp and the FAR. In the current configuration, four qualifiers for supernova alerts have been defined:
\begin{itemize}
    \item \texttt{internal}: alert stream only distributed internally for monitoring purposes, with an hourly FAR;
    \item \texttt{snews-low}: alerts with a FAR below one per day; sent to SNEWS for testing purposes and internally exploited for the routine testing of the quasi-online analysis;
    \item \texttt{snews}: alerts with a FAR below one every eight days; sent to SNEWS for alerting purposes.
    \item \texttt{high-Z}: alerts with a FAR below one per year, aimed at public advertising via the GCN.
\end{itemize}

 Since consecutive SN Z messages are based on overlapping time intervals, the generation of multiple alerts from a single fluctuation of the number of SN events must be prevented. To this purpose, a dead time interval of one second is applied at every generated alert to the corresponding FAR threshold. As a consequence, the alerts will have an \emph{effective FAR} lower than the threshold value by a factor up to ten.
 
 A handler application is designated to read the alert messages from the dispatcher and, for the \texttt{snews} alert qualifier, sends an alert datagram to the SNEWS network. The communication with the SNEWS system is based on a simple proprietary protocol based on TCP \cite{RFC0793} socket communication. The low threshold \texttt{snews-low} alerts are sent to a separate coincidence server, mantained by SNEWS for testing and development purposes. The alert and the corresponding transmission logs are stored in a dedicated database. Typically, the alert message is sent in less than 20~s from the generation of the corresponding data offshore. This makes KM3NeT one of the fastest alert sending experiments among the neutrino detectors participating in SNEWS, together with NO$\nu$A and HALO~\cite{SNEWS2.0}. 

 In order to initiate the quasi-online analysis, the alert messages are propagated from the multi-messenger dispatcher to the local dispatcher of each detector (see \secref{s:daq}). To trigger the dump of the buffered data, a dedicated application running at each detector site translates the selected internal alerts into a control message sent to the data acquisition system. The final results of the quasi-online analysis ($T_0$, neutrino light-curve, refined significance) can be shared with external networks. The principal destination for this is the SNEWS 2.0 system currently under development. The data exchange with SNEWS 2.0 is managed by the SNEWS 2.0 handler (see \figref{fig:quasi-online}) and will rely on the HOPSKOTCH publish-subscribe message system developed within the SCIMMA program \cite{baxter2021agile}.

\section{Triggered follow-up searches}
\label{s:triggered_search} 
 KM3NeT implements a triggered analysis strategy for the follow-up of external multi-messenger alerts. This aims to provide a fast response following a semi-automated analysis. The events of interest include unmodelled gravitational-wave burst signals \cite{Roma:2019kcd}, very close gamma-ray or fast radio bursts, or other types of astrophysical transients expected to produce \SIrange{1}{100}{MeV} neutrinos.  Ultimately, a confirmed CCSN detection, signaled by a public alert provided by SNEWS or individual experiments (such as the SuperKamiokande \cite{Ikeda:2007sa} public alert stream \footnote{\url{https://gcn.gsfc.nasa.gov/sk_sn.html}}), would represent a golden trigger for this search. The expected latency of the triggered analysis is less than one minute.
 
 \subsection{External alert reception}

 Multi-messenger alerts are routinely exchanged via worldwide networks. Individual observatories can subscribe for the reception of human- and machine-readable alerts through a number of channels. In KM3NeT, alerts coming from different networks are automatically collected and distributed to the multi-messenger dispatcher by an alert broker module.

 The two main sources of alerts are the GCN and the SNEWS system itself. The GCN is mainly used to distribute the alert messages produced by the gamma-ray burst satellites (Swift, Fermi-GBM, INTEGRAL), the gravitational-wave interferometers (LIGO, VIRGO, KAGRA) and some neutrino detectors (IceCube, ANTARES). SNEWS is also distributing their neutrino triggers via GCN, but allows in parallel individual experiment to establish a socket-based direct channel. To date, SNEWS has never sent a CCSN alert since its inception, while test alerts are produced once per week.

 The GCN alert messages (\emph{notices}) are received in a VOEvent format \cite{Seaman:2011cb}. Each notice contains a unique ID, the observing role (observation or test), the alert and the notice UTC times, its eventual location and its significance. The alert broker module assures the connection to the GCN server, the selection and the parsing of the notices, and a log of the notice information. The selected triggers are translated into a JSON message and distributed to the multi-messenger dispatcher. This allows to rapidly look into the KM3NeT data for the identification of a transient that does not pass the KM3NeT significance thresholds for alerts. The information of each selected external trigger and the results of the triggered analysis are stored in the alert database.
 
 \subsection{Triggered analysis}
 
 The follow-up of an astrophysical transient candidate is performed using the output of the real-time analysis described in \secref{s:rta}. To this purpose, in parallel with the processing by the KM3NeT CCSN trigger, the SN summary data received by the multi-messenger dispatcher from the two detectors are collected and stored in a dedicated time series database. On the reception of an alert, a query of the time series database is performed. In the typical case of a CCSN search, the $\SI{500}{ms}$ time window that has the starting time nearest to the time of the trigger is considered for evaluating the detection significance. The time window can be adjusted to match the time scale of different astrophysical transients. The quality score and the significance combination logic, described in \secref{s:pipeline} for the CCSN trigger, are applied in a similar manner to this analysis. 
 
 For each alert follow-up, a summary message is produced reporting the evaluated combined significance, with the number of observed events and the background expectation in each detector. If no significant excess is observed, upper limits on the signal expectation value are automatically derived following the Feldman-Cousins \cite{Feldman:1997qc,Cousins:2007hgg} statistical approach.
 To this end, the two detectors are treated as a single experiment, summing the background expectations and the number of events detected by ARCA and ORCA. Following common conventions, the combined upper limit on the number of expected signal events at a 90\% confidence level (CL), $\mu_{s,0.9}$, is estimated.
 
 Considering one or more reference neutrino spectra for which the detector effective mass is known, it is possible to constrain the signal scale, defined as the ratio between the total energy released by the source and its squared distance. From the signal expectation $\mu_s(E_0, d_0)$ estimated from a total energy $E_0$ and a distance $d_0$, the number of signal events scales naturally as:
 \begin{equation}
     \mu_s(E,d) =  \mu_s(E_0, d_0) \, \frac{E}{E_0} \, \left( \frac{d_0}{d} \right)^2 
 \end{equation}
 so the derived 90\% upper limit, $\mu_{s,0.9}$, is translated into an upper limit on the signal scale $E\,d^{-2}$ as follows:
 \begin{equation}
     \frac{E}{d^2} \leq \left( \frac{E}{d^2} \right)_{0.9} = \frac{E_0}{d_0^2}  \frac{\mu_{s,0.9}}{\mu_s(E_0,d_0)}
 \end{equation}
 An upper limit on the signal expectation defines a parabola in the $(d,E)$ plane, $E = A \, d^2$, where $A$ is a constant dependent on the detector effective mass for the given spectral properties, ($\avg{E}$, $\alpha$), of the flux. The area above the parabola represents the parameter space that can be excluded at a 90\% CL. For the case of a power-law spectrum, a similar scaling can be defined by replacing the total energy, $E$, with the flux normalisation $\Phi$, and estimating the expected number of signal events $\mu_s(\Phi_0,d_0)$ for one or more given values of the spectral index.
 
 The time profile of the number of SN events around the time of the search is also evaluated, in order to facilitate the subsequent manual check.
 
 A report following the first realisation of the follow-up analysis is given in \secref{s:gw-followup}.

\section{Current status, perspectives and first results}
\label{s:current}

KM3NeT has been operating the first detection units of the ARCA detector since 2016 followed by those of ORCA since 2017. ORCA has been continuously operating with four detection units since mid-2019 and has grown to six detection units in early 2020. A total of 216 optical modules are currently deployed. The equivalent effective mass for the 6--11 multiplicity selection is \SIrange{0.14}{0.28}{kton}, depending on the considered spectrum among the three progenitors \cite{KM3NeT2021-CCSN}. The corresponding effective mass for the all-coincidence sample used in the quasi-online analysis is \SIrange{4.1}{7.0}{kton}. In 2019, ARCA has been operated with one detection unit, before undergoing a switch-off period for the refurbishing of the shore station. Data taking has resumed in the last months of 2020. The first implementation of the KM3NeT CCSN real-time trigger has been activated in the first months of 2019. Since then, the CCSN real-time analysis has been continuously running, with progressive revisions and improvements of the software. In spring 2021, new detector elements were deployed, bringing ARCA to a size of six detection units. This section outlines the current triggering capabilities and the first multi-messenger follow-ups conducted in this time period.

\subsection{Alert horizon}

\begin{figure}
    \centering
    \includegraphics[width=0.45\textwidth]{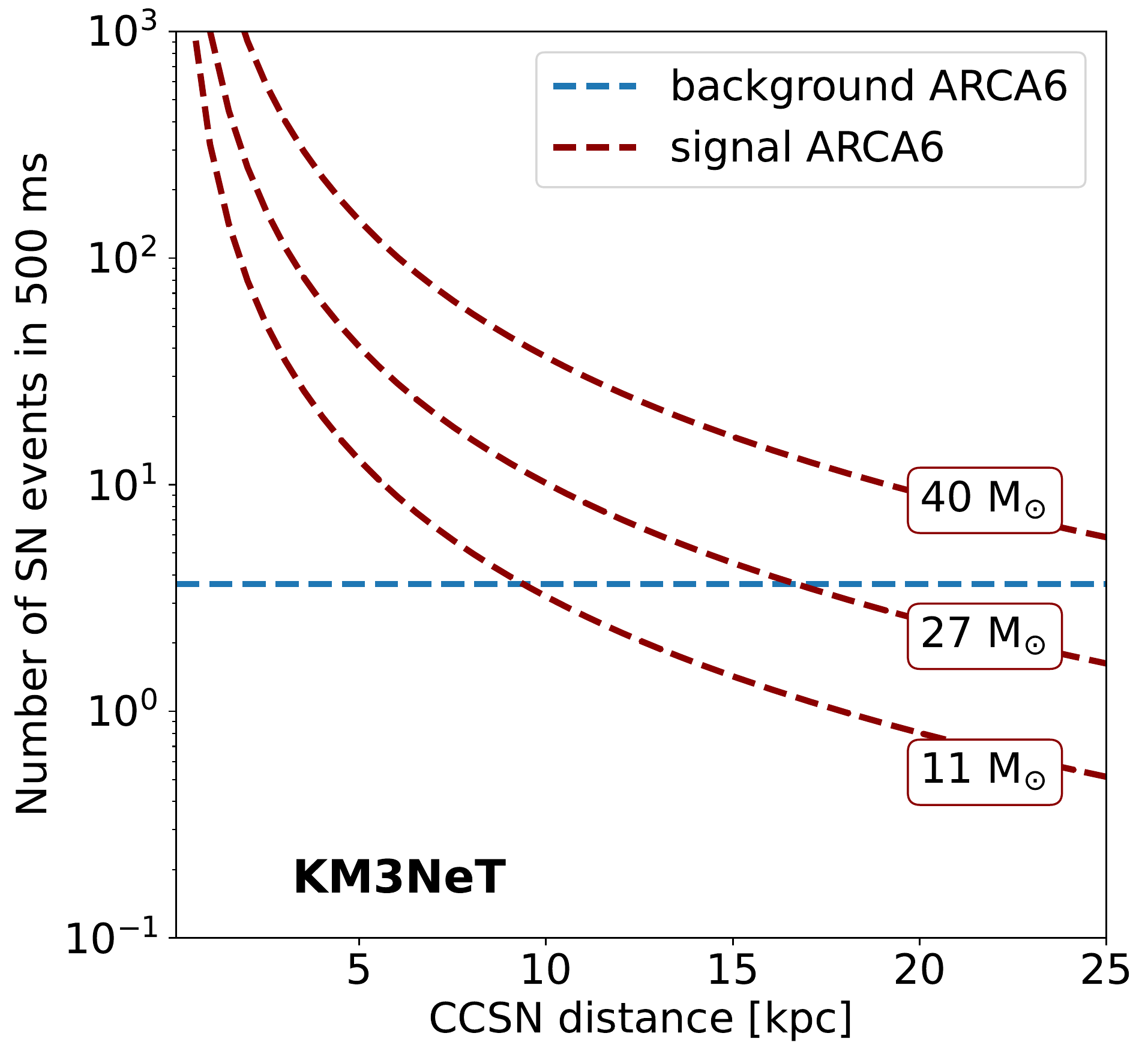}%
    \\
    \includegraphics[width=0.45\textwidth]{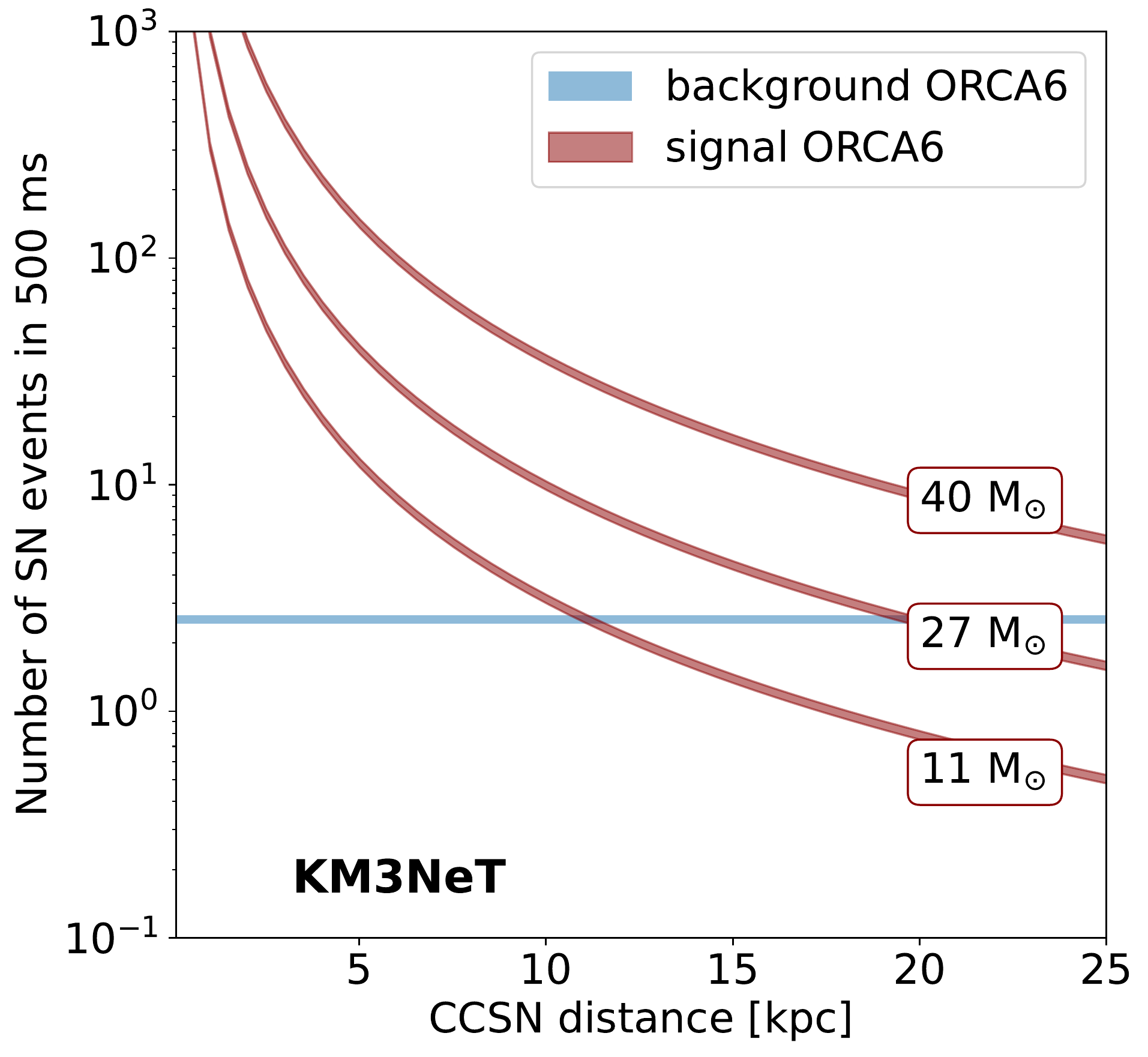}%
    \caption{Expected number of signal events in the ARCA6 (top) and ORCA6 (bottom) detectors compared to their background expectations. The $\SI{27}{\solarmass}$ model considered in \litcite{KM3NeT2021-CCSN} is taken as a benchmark. The average photon detection efficiency of the PMTs of the two detectors is taken into account to determine the signal expectation. The instrumentation efficiency is accounted for by considering its variation within the interval corresponding to a cumulative livetime fraction of 68\% (see \figref{fig:efficiency-parameterisation}). Its effect is reflected on the signal and background expectations of ORCA6 (red and blue bands), while it is negligible for ARCA6 (red and blue dashed lines).}
    \label{fig:horizon1}
\end{figure}

\begin{figure}
    \includegraphics[width=0.45\textwidth]{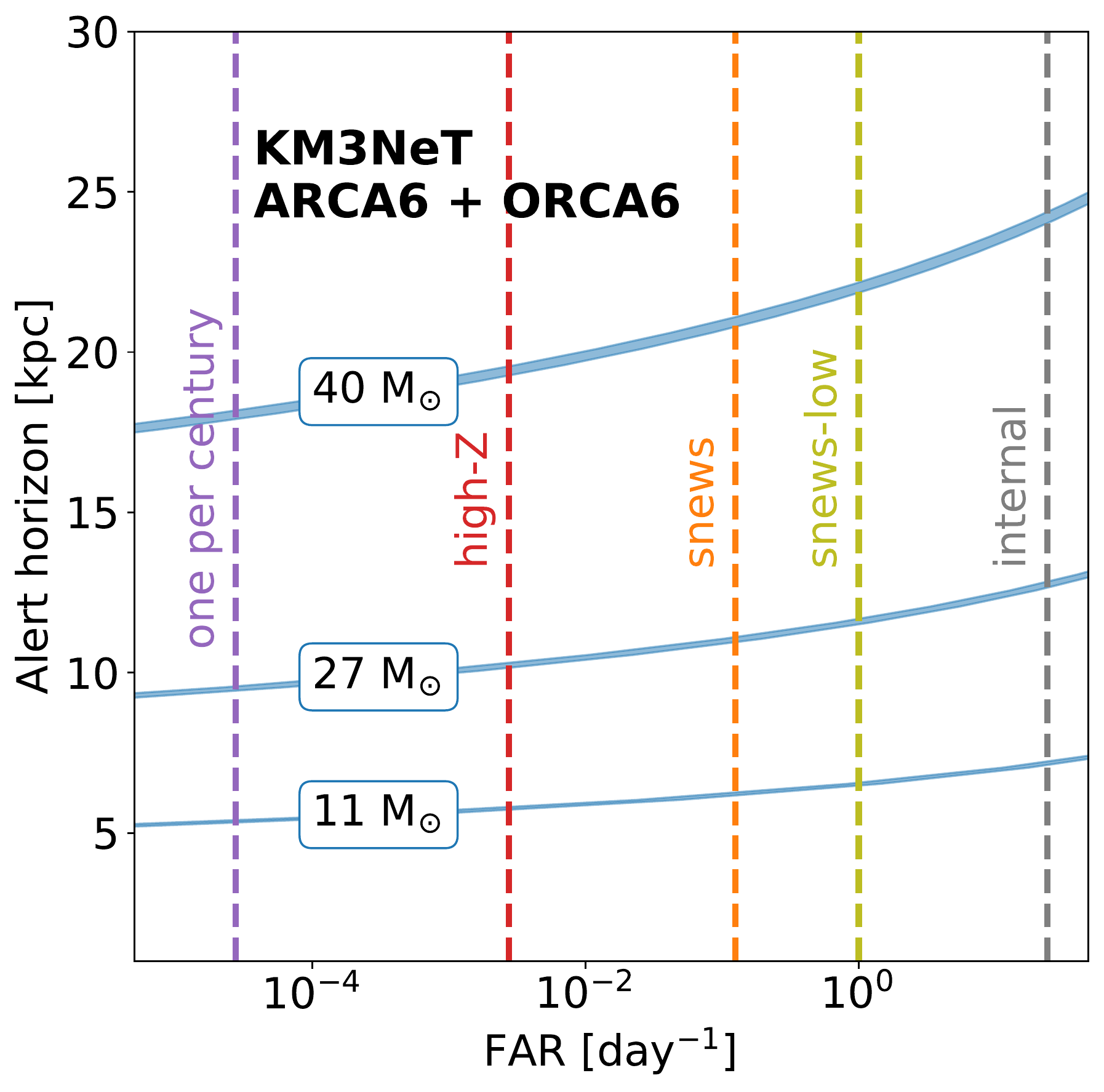}%
\caption{
    Maximal source distance (blue band) for which the ARCA6 + ORCA6 detector is able to produce an alert for the FAR thresholds corresponding to the qualifiers defined in \secref{s:alerts} plus an additional threshold corresponding to a FAR of one per century (vertical dashed lines). The band width accounts for the variability of the background and signal expectations shown in \figref{fig:horizon1}.}
    \label{fig:horizon2}
\end{figure}

 The triggering capability of KM3NeT is given for a currently operational setup of six ARCA and six ORCA detection units (ARCA6 and ORCA6). For a source at 10~kpc, the number of SN events in the 6--11 multiplicity range expected from the $\SI{11}{\solarmass}$, $\SI{27}{\solarmass}$ and $\SI{40}{\solarmass}$ progenitors is of 0.63, 2.0 and 7.2 per detection unit (18 DOMs) respectively. The expected number of background events in a 500 ms time window is estimated from archival data to be of $\sim 3.5$ in ARCA6 and $\sim 2.8$ in ORCA6. In \figref{fig:horizon1}, the expected number of signal events in ARCA6 (top) and ORCA6 (bottom) is shown as a function of the distance to the source for the three progenitors, and compared to the background expectation values. The difference in the background rates between the two detectors comes from the higher efficiency of the muon rejection in ORCA, achieved thanks to the smaller distance between its optical modules~\cite{KM3NeT2021-CCSN}.
 The effective conditions of the detectors operation are accounted for, including the number of functional DOMs, the average photon detection efficiency of the PMTs~\cite{KM3NeT:2019-MuonDepth} and the variability of the instrumentation efficiency. From the signal and background expectation values, the combined significance expected from a KM3NeT observation is calculated. The FAR is determined by multiplying the equivalent p-value by the update frequency ($\SI{10}{Hz}$) of the trigger time window, and represented in units of one per year. As described in \secref{s:alerts}, the effective FAR of the alerts for a given threshold is systematically lower than the threshold value, thus some sensitivity is traded in favour of robustness. In \figref{fig:horizon2}, the maximum triggering distance (alert horizon) for the three evaluated stellar progenitors is reported as a function of the FAR threshold for the combination of ARCA6 and ORCA6. The alert thresholds defined in \secref{s:alerts} are indicated. An additional line is used to mark a FAR equal to one-per-century, equivalent to the one expected from the public SNEWS alerts. 
 
 When considering the scenario of a \SI{27}{\solarmass} progenitor, KM3NeT in its current configuration is able to send an alert to SNEWS up to almost 11~kpc. A FAR as low as one per century is reached at a distance of about 9~kpc. This will improve with the installation of new detection units, as the construction of the detector continues.

\begin{figure}
    \centering
    \includegraphics[width=0.45\textwidth]{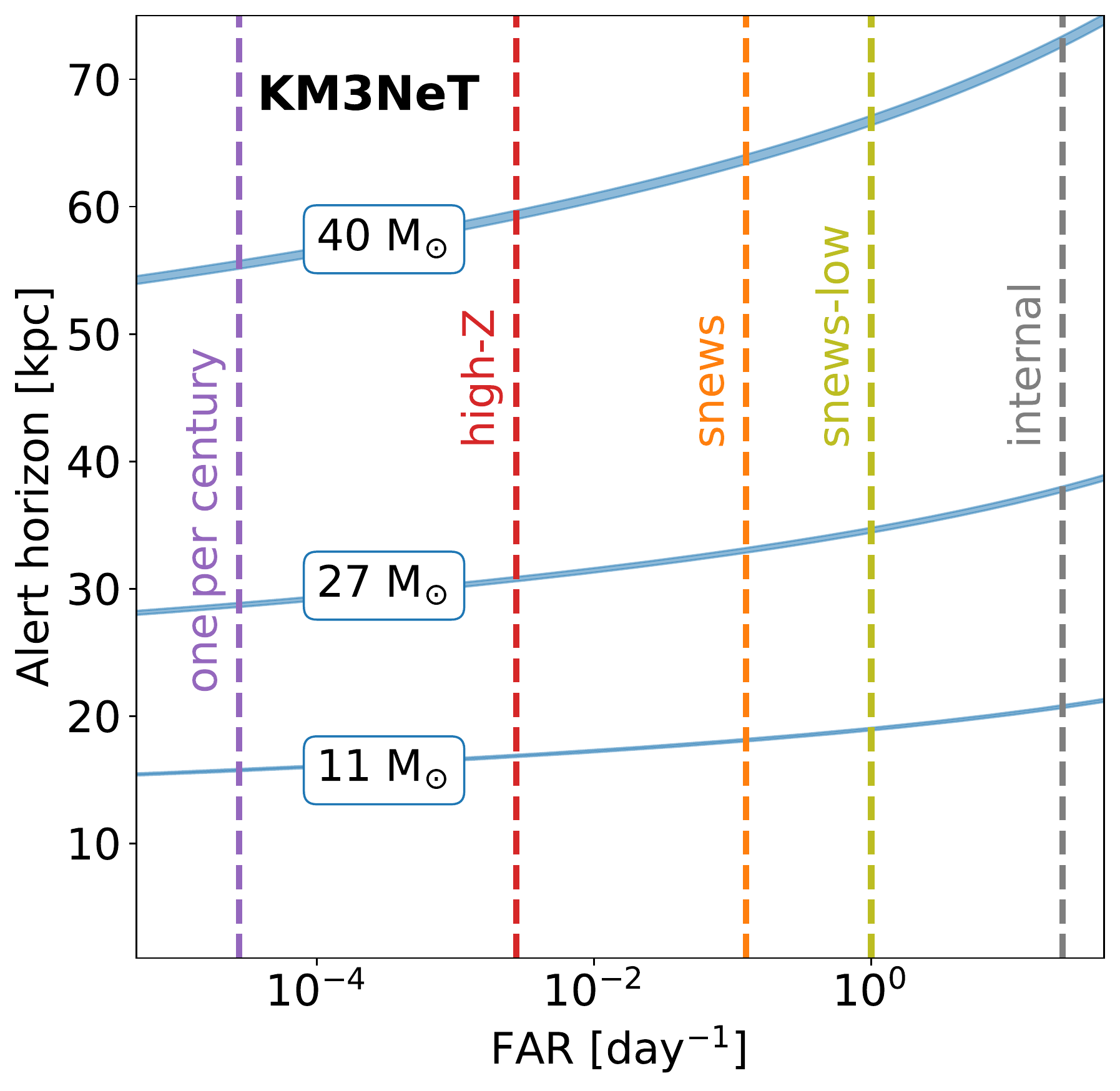}
    \caption{Maximal source distance (blue band) for which the complete KM3NeT ARCA + ORCA detector will be able to produce an alert for the FAR thresholds corresponding to the qualifiers defined in \secref{s:alerts} plus an additional threshold corresponding to a FAR of one per century (vertical dashed lines). The band width accounts for the variability of the background and signal expectations, assuming the same relative variations of \figref{fig:horizon2}.}
    \label{fig:horizon-km3net}
\end{figure}

The projected performance for the completed KM3NeT detector, comprising 230 detection units for ARCA and 115 detection units for ORCA for a total of $\sim 6000$ optical modules ($\sim \num{200000}$ PMTs), is reported in \figref{fig:horizon-km3net}. In this final configuration, ARCA and ORCA will have a combined effective mass of 2.0--4.4~kton for the 7--11 multiplicity selection and of 120--200 kton when considering all coincidences. For the scenario of a $\SI{27}{\solarmass}$ progenitor, KM3NeT will be essentially able to produce alerts for a CCSN occurring up to the edge of the Milky Way. For the more conservative case of the $\SI{11}{\solarmass}$ progenitor, the alert horizon will be beyond the Galactic Centre ($\sim \SI{10}{kpc}$) for all the alert thresholds. For the $\SI{40}{\solarmass}$ black-hole forming scenario, the SNEWS alert horizon extends beyond $\SI{60}{kpc}$ ensuring coverage up to the Large Magellanic Cloud.

\subsection{Follow-up of gravitational-wave alerts}
 \label{s:gw-followup}
 Core-collapse supernovae are candidate sources of gravitational waves. In some scenarios, both neutrinos and gravitational waves are predicted in correspondence to the core collapse~\cite{MM_CCSN}. The LIGO and Virgo gravitational-wave detectors implement a dedicated search for unmodelled burst-like signatures from sources different from typical compact binary mergers \cite{LIGOScientific:2021hoh}.

 Binary neutron stars are also thought to produce thermal neutrinos in the MeV energy regime together with their well-known gravitational-wave signature at the time of the merger~\cite{BNS_MeVnu}. However, such mergers are very rare in our Galaxy, with an estimated rate of few tens per million years~\cite{Pol:2020tfz}, and current and near-future neutrino detectors will only be sensitive to close-by sources ($d \lesssim \SI{1}{\mega\parsec}$).

 The first gravitational-wave alerts from the LIGO-Virgo unmodelled search were published during the observing run designated as O3\footnote{\url{https://gracedb.ligo.org/superevents/public/O3/}}. The two events are designated S191110af and S200114f. Follow-ups were performed in KM3NeT with the ORCA4 detector operating at that time. The results have been reported in two corresponding GCN circulars \cite{GCN26249,GCN26751}. While the procedure was based on offline data, it represents a prototype case for the triggered search described in \secref{s:triggered_search}. Given the later retraction of S191110af, the results are presented and discussed only for the event S200114f. 
 
 The gravitational wave event S200114f was identified on the 14th of January 2020 by the unmodelled LIGO-Virgo pipeline \cite{GCN26734}. The probability of this event originating from background is of about one in 25 years. Moreover, the event was well localised ($\sim \SI{400}{deg^2}$ 90\% confidence level area), and compatible with the Galactic Plane location. For these reasons, the event was of great interest as a potential MeV neutrino emitter. The very near star Betelgeuse ($\alpha$ Ori), known to be a CCSN progenitor candidate, is located in the same region of the sky, although outside of the 90\% confidence level area.

\begin{figure*}[!ht]
    \centering
    \includegraphics[width=0.8\textwidth]{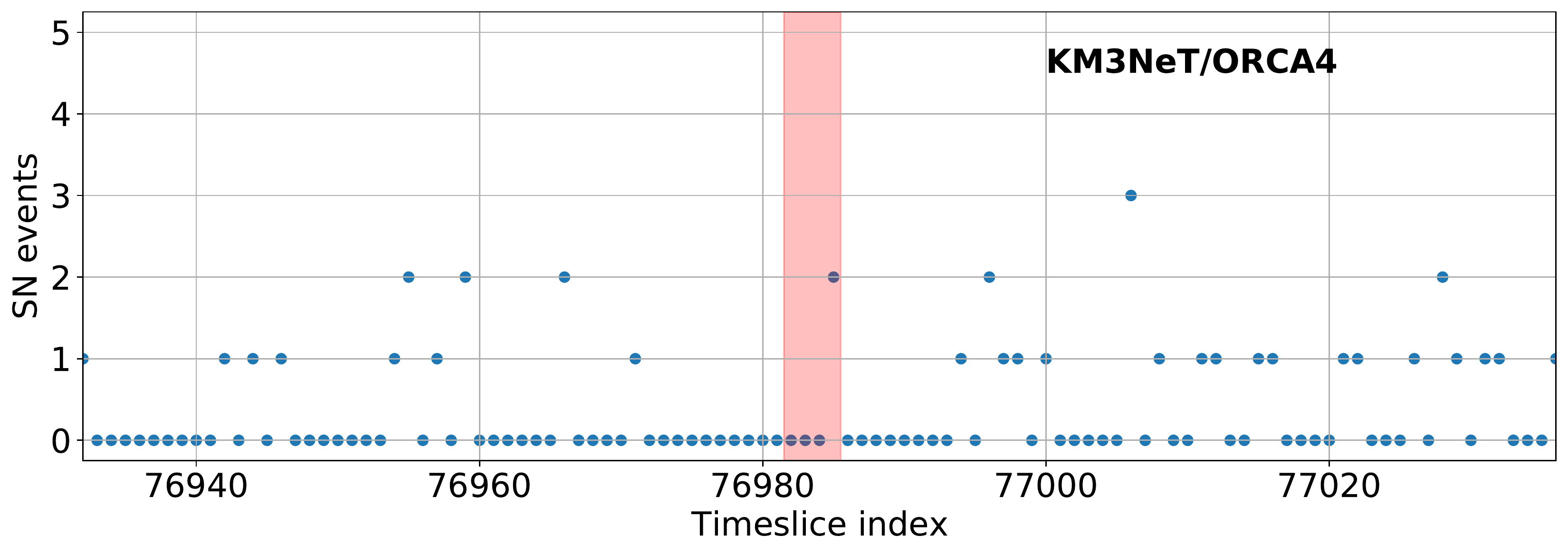}
    \caption{Number of SN events per timeslice as a function of time (blue markers) for the 10 seconds around the event time (T0) of the LIGO-Virgo gravitational wave alert S200114f. The timeslices corresponding to the search time window are highlighted in red. The first timeslice in the search time window starts 39 ms before T0.}
    \label{fig:followup_data}
 \end{figure*}

 The ORCA4 data were used to search for an excess over the background expectation at the time of the gravitational-wave detection. The analysis was performed over a time window of 400~ms from the GW trigger time. The use of a 400~ms window has been since superceded by the 500~ms considered in this paper, after the final optimisation of the offline analysis described in Ref.~\cite{KM3NeT2021-CCSN}. Two SN events were found in the data, while 1.45 were expected from background, after correcting for the estimated instrumentation efficiency. The observation is not significant, having a p-value of 0.4. The 90\% confidence level upper limit on the number of signal events was evaluated using the Feldman and Cousins approach. The time profile of the SN events around the search window is shown in Figure~\ref{fig:followup_data}. 
 
 The information about the GW event and the results of the follow-up are summarised in Table~\ref{t:event_followup}. The absence of a detection allowed to set constraints on the possible CCSN origin of the gravitational wave signal. Lower limits on the distance to the source were determined. With only four detection units in operation (while 3 blocks of 115 will be available in the complete detector configuration), KM3NeT was able to rule out a CCSN source closer than 6--12~kpc for the typical flux assumptions of the analysis \cite{KM3NeT2021-CCSN}. In the scenario of a failed CCSN with black hole formation, the possibility of a core-collapse event could be rejected up to a distance of $\sim\SI{21}{\kilo\parsec}$. The results are summarised in Table~\ref{t:distlim}.
 
\begin{figure}
     \centering
     \includegraphics[width=\linewidth]{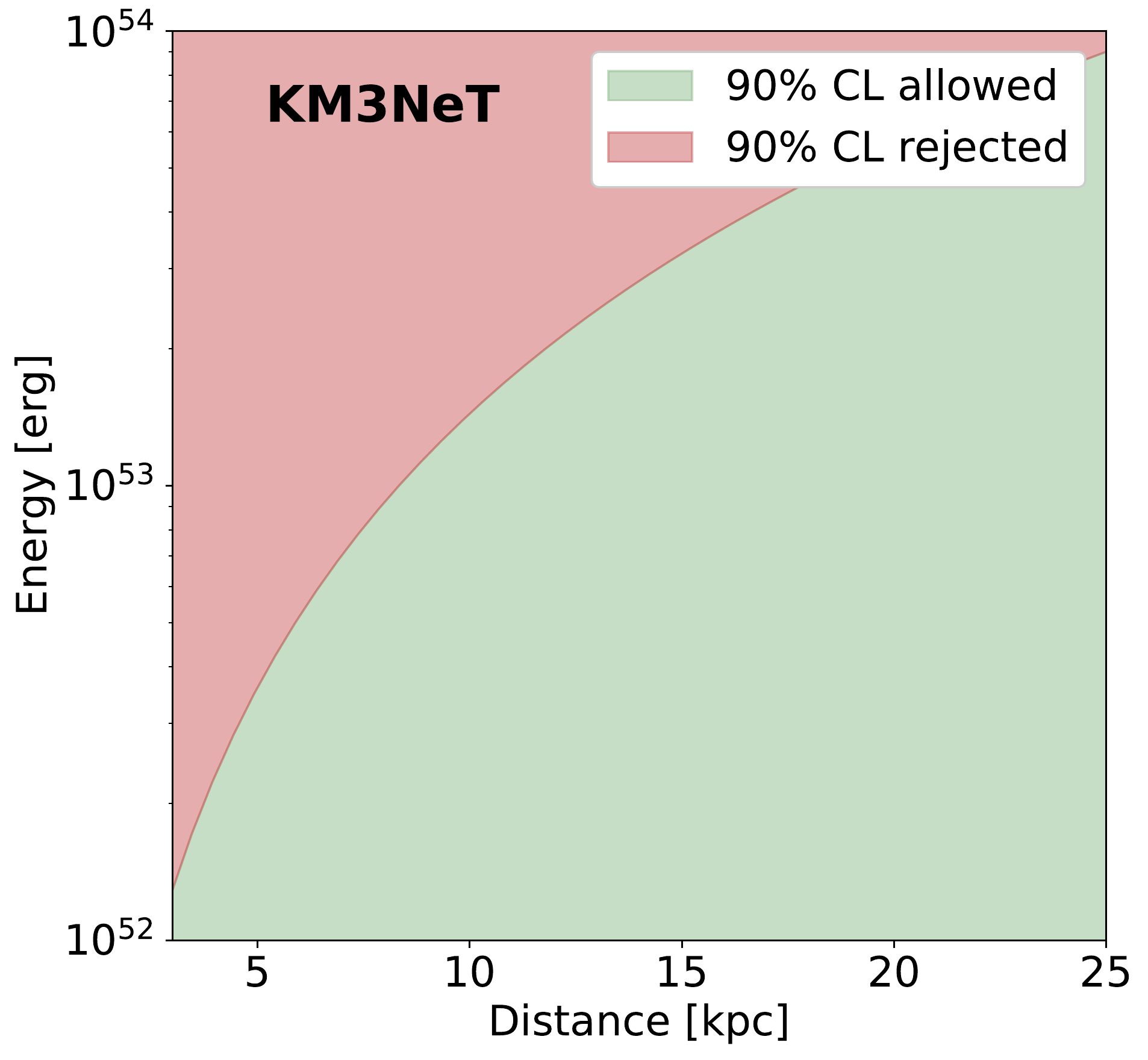}
     \caption{Exclusion region at 90\% CL for total released energy versus CCSN candidate distance for the followup of S200114f, determined by an upper limit of $\mu_{s,0.9} = 4.8$ against an expectation value of $\mu_{s} \simeq 5$ for the considered spectral parameters at a distance to the progenitor of 10 kpc.}
     \label{fig:sn-ed2-limit}
 \end{figure}

 Following the protocol described in \secref{s:triggered_search}, assuming a typical quasi-thermal spectrum \cite{Keil:2002in} with parameters $\avg{E_{\nu}} = \SI{15}{MeV}$ and $\alpha = 3$, a combined total energy and distance limit $E\,d^{-2} \lesssim (\SI{3E53}{\erg})\, (\SI{10}{\kilo\parsec})^{-2}$ is derived. The corresponding exclusion region is shown in \figref{fig:sn-ed2-limit}.
 
\begin{table}
\caption{S200114f follow-up summary and results.}
\begin{center}
 \begin{tabular}{@{}lc@{}}
 \toprule
 Event                        & S200114f \\ \midrule
 T0 date                      & 2020--01--14   \\
 T0 time                      & 02:13:41.239 \\
 SN events {[}T0 - 39ms, T0 + 361 ms{]} & 2        \\
 Instrumentation efficiency $\eta$   & 0.875  \\
 Expected background $\mu_b(\eta)$   & 1.45   \\ 
 Feldman-Cousins $\mu_{s,0.9}$           & 4.8     \\ \bottomrule
 \end{tabular}%
\end{center}
\label{t:event_followup}
\end{table}

\begin{table}[!ht]
\caption{Lower limits on the CCSN distance at the 90\% confidence level ($d_{0.9}$) for the three progenitors considered in Ref.~\cite{KM3NeT2021-CCSN}.}
\begin{center}
\begin{tabular}{ccccc}\toprule
{\bf Progenitor } & {\bf $\mathbf{d_{0.9}}$ [kpc]}  \\ \midrule
11 M$_{\odot}$ & 6 \\ 
27 M$_{\odot}$ & 11.5 \\ 
40 M$_{\odot}$ & 21 & \\ \bottomrule 
  \end{tabular}
\label{t:distlim}
\end{center}
\end{table}

\section{Conclusions}
\label{s:conclusion}

An analysis pipeline for the search of a core-collapse supernova neutrino signal in KM3NeT has been developed and implemented as a real-time system. It has been operational since mid-2019, continuously processing the data acquired by the first detection units deployed in the sea. The real-time trigger is based on the event selection introduced for the estimation of the KM3NeT sensitivity~\cite{KM3NeT2021-CCSN}. In addition, a buffering of low-level data exploitable in the follow-up of confirmed detections has been introduced, together with methods for a quasi-online astronomy analysis of the detected neutrino light curve. A brokering strategy to manage internal and external alerts has been implemented.
The continuous evaluation of the time-dependent detection significance of the combined ARCA and ORCA signals allows to generate internal alerts. Corresponding false alarm rate thresholds are defined for the triggering of the quasi-online analysis and the sending of alerts to external networks. In particular, KM3NeT is taking part in the Supernova Neutrino Early Warning System (SNEWS), being one of the detectors with the lowest latencies (about 20 s). With the current detector configuration (ARCA6 + ORCA6), the horizon up to which a significant detection can be expected is around 10~kpc.

The quasi-online capabilities allow KM3NeT to share the light curve data together with a determination of the time of arrival of the neutrino burst within a few minutes from the detection. This is of fundamental importance for the triangulation and advanced time-domain analyses for KM3NeT and the SNEWS 2.0 system. KM3NeT is currently a key player in the development and testing of the SNEWS 2.0 infrastructure. 

 Finally, external alerts received via the GCN system or dedicated channels are followed up with a search in the archival data of the supernova trigger, allowing for a fast determination of the significance and the calculation of upper limits in the absence of a signal.

 This analysis has also been used to search for neutrino counterparts of two gravitational-wave unmodelled candidate events sent by the LIGO-VIRGO Collaboration during the O3 run. No signal has been found and constraints on the presence of CCSNe have been set. These results have been reported in the first two GCN circulars issued by the KM3NeT Collaboration \cite{GCN26249,GCN26751}.

 The operation of the CCSN real-time search pipeline and the first follow up of multi-messenger alerts described in this paper are the first activities of the KM3NeT Collaboration in the domain of multi-messenger astronomy.

\begin{acknowledgements}
The authors acknowledge the financial support of the funding agencies: Agence Nationale de la Recherche (contract ANR-15-CE31-0020), Centre National de la Recherche Scientifique (CNRS), Commission Europ\'eenne (FEDER fund and Marie Curie Program), Institut Universitaire de France (IUF), LabEx UnivEarthS (ANR-10-LABX-0023 and ANR-18-IDEX-0001), Paris \^Ile-de-France Region, France; Shota Rustaveli National Science Foundation of Georgia (SRNSFG, FR-18-1268), Georgia; Deutsche Forschungsgemeinschaft (DFG), Germany; The General Secretariat of Research and Technology (GSRT), Greece; Istituto Nazionale di Fisica Nucleare (INFN), Ministero dell'Universit\`a e della Ricerca (MIUR), PRIN 2017 program (Grant NAT-NET 2017W4HA7S) Italy; Ministry of Higher Education Scientific Research and Professional Training, ICTP through Grant AF-13, Morocco; Nederlandse organisatie voor Wetenschappelijk Onderzoek (NWO), the Netherlands; The National Science Centre, Poland (2015/18/E/ST2/00758); National Authority for Scientific Research (ANCS), Romania; Ministerio de Ciencia, Innovaci\'{o}n, Investigaci\'{o}n y Universidades (MCIU): Programa Estatal de Generaci\'{o}n de Conocimiento (refs. PGC2018-096663-B-C41, -A-C42, -B-C43, -B-C44) (MCIU/FEDER), Generalitat Valenciana: Prometeo (PROMETEO/2020/019), Grisol\'{i}a (ref. GRISOLIA/2018/119) and GenT (refs. CIDEGENT/2018/034, /2019/043, /2020/049) programs, Junta de Andaluc\'{i}a (ref. A-FQM-053-UGR18), La Caixa Foundation (ref. LCF/BQ/IN17/11620019), EU: MSC program (ref. 101025085), Spain. 
\end{acknowledgements}

\bibliographystyle{hunsrt} 
\bibliography{references,references_unpublished_arxiv}

\end{document}